\documentstyle[12pt,epsf]{article}

\makeatletter

\def\section{\@startsection {section}{1}{\z@}{24pt plus 2pt minus 2pt}
{12pt plus 2pt minus 2pt}{\large\bf}}

\def\subsection{\@startsection {subsection}{2}{\z@}{12pt plus 2pt minus 2pt}
{12pt plus 2pt minus 2pt}{\normalsize\bf}}
\makeatother

\begin{document}

%don't want date printed
\date{}

\title{\Large\bf On the Origin of Cusps in Dark Matter Halos}

\author{Toshiyuki Fukushige and Junichiro Makino\\
Department of General Systems Studies,\\
College of Arts and Sciences, University of Tokyo,\\
3-8-1 Komaba, Meguro-ku, Tokyo 153, Japan\\
Email: fukushig@chianti.c.u-tokyo.ac.jp}

\maketitle

\begin{abstract}

Observed cusps with density profiles $\rho\propto r^{-1}$ or shallower,
in the central regions of galaxies, cannot be reproduced in the standard
Cold Dark Matter (CDM) picture of hierarchical clustering.  Previous
claims to the contrary were based on simulations with relatively few
particles, and substantial softening.  We present simulations with
particle numbers an order of magnitude higher, and essentially no
softening, and show that typical central density profiles are clearly
steeper than $\rho\propto r^{-1}$.  The observed shallower profiles
may have formed through the smoothing effect of the spiral-in of
central black holes in previous merger phases.  In addition, we
confirm the presence of a temperature inversion in the inner 5 kpc of
massive galactic halos, and illustrate its formation as a natural result of the
merging of unequal progenitors.

\end{abstract}

%\keywords{cosmology:theory --- dark matter --- galaxies:kinematics and
%dynamics --- galaxies:formation --- method: numerical}

\section{Introduction}

Recent high resolution ground-based and Hubble Space Telescope
observations (Lauer et al.  1995) have revealed that elliptical
galaxies do not have a constant-density core and the densities
continue to rise until the resolution limit.  In faint ellipticals,
the density profile at the central region increases roughly as
$\rho\propto r^{-2}$, and in bright ellipticals it increases as $\rho
\propto r^{-1}$ or shallower (Merritt and Friedman 1996). 

Several simulations demonstrated that the dark matter halos formed
through hierarchical clustering were not well approximated by
isothermal spheres, and better fitted by a Hernquist model (Hernquist
1990) or similar models with $\rho \sim 1/r$ at the center (Dubinski
and Carlberg 1991; Navarro, Frenk, and White 1996a, 1996b).  In these
simulations, halos have a power law index which changes from around
$-1$ to $-3$ or $-4$ as the radius increases.  Recently, Moore (1994)
and Flores and Primack (1995) have argued that the gently rising
$\rho(r)$ curves of dwarf galaxies are inconsistent with a $r^{-1}$
central cusp, and therefore challenge the hypothesis of CDM halos.

However, whether the results of numerical simulations are physically
valid or are numerical artifacts remains unclear for the following
reasons.  First, no physical mechanism has been presented so far for
the formation of cusps found in simulations (White 1996).  Secondly,
since the mass resolution in these simulation was rather low, the
central structure may have been affected by two-body relaxation
effects (Quinlan 1996, Steinmetz and White 1996, Fukushige and Makino
1996).  Thirdly, the central structures of halos formed in these
simulations are strongly affected by the potential softening used, and
show rapid change in the power index of $\rho(r)$ around a radius not
much larger than the potential softening length. 

In this letter, we have been able to separate physically real effects
from numerical artifacts, through $N$-body simulations of hierarchical
clustering with a resolution far higher than those in previous work.
The number of particles we used ($N=786,400$) is more than 10 times
larger than those used in previous simulations ($N\sim
10,000-30,000$), and our softening is significantly smaller (\S2).

Our results show that dark halos formed through hierarchical
clustering typically exhibit an inwardly decreasing temperature
structure in their inner regions, with a density cusp shallower than
$\rho\propto r^{-2}$, but steeper than $\rho\propto r^{-1}$ (\S3).  We
illustrate the origin of this behavior through an idealized merger
simulation of different types of galaxies (\S4), which enables us to
give a physical explanation of temperature inversion (\S5).

\section{Method}

Initial conditions were constructed following 
Dubinski and Carlberg (1991).  We assigned initial positions and
velocities to particles in a spherical region with a radius of 2Mpc
surrounding a density peak selected from a discrete realization of a
standard CDM model ($H_o = 50$km/s/Mpc and $\Omega = 1$).  The peak
was chosen from a realization of the density contrast in an 8Mpc box,
using COBE normalization. 
The peaks were found by smoothing the density with a Gaussian filter
of radius 0.75 Mpc.  

We followed the evolution of a density peak through direct $N$-body
simulation. We added the local Hubble flow and integrated the orbits
directly in physical space, with $N=786,400$, an individual particle
mass of $4.0 \times 10^6M_{\odot}$, and a Plummer softened potential
with a length of $\varepsilon=0.14$kpc.  We started the simulation at
$z\sim 46$.   We did not include tidal effects from outside our
2Mpc sphere, since we are mainly interested in the core
properties within 10kpc, where tidal effects from outside 2Mpc
scale are negligible.

We used a 4-th order Hermite integration scheme (Makino and
Aarseth 1992) with individual (hierarchical) timestep algorithm
(McMillan 1986, Makino 1991a).

For the force calculation, we used the GRAPE-4 (Taiji et al. 1996), a
special-purpose computer designed to accelerate $N$-body simulations
using a Hermite integrator and a hierarchical timestep algorithm.  The
total system consists of 1692 pipeline processor chips dedicated to
gravity calculation and has a peak performance of 1 Tflops.  The
calculation with $N=786,400$ took about 180 CPU hours ($5.7\times
10^9$ particle steps) using 3/4 of the total system. The sustained
speed of computation was 406 Gflops.  Since the force calculations on
GRAPE-4 are effectively of double-precision accuracy (Makino et
al. 1996), our simulations exhibit  much higher accuracy both for force
calculations and orbit integrations than  previous simulations.

\section{Hierarchical Clustering Simulations}

Figure 1 shows the particle distributions in our simulation at the
redshift $z=$(a)$8.7$, (b)$5.1$.and (c)$1.8$.  Figure 2 shows the
density and temperature structure of the halo at $z=1.8$, well after
most of the mergings has already taken place (results at $z=0$ are
similar, but these earlier stages shows even less effects of two-body
relaxation).  In figure 2 we can see a clear temperature decrease
toward the center within 5kpc. This non-isothermal structure produces
a density cusp shallower than $\rho\propto r^{-2}$.  In our
simulation, the structure outside $1$ kpc is not affected by two-body
relaxation after the formation of a large single halo at $z\sim 3$; only
the central region within $1$ kpc shows some expansion because of
two-body relaxation effects. 

The large potential softening employed in previous studies has
produced spurious structures, which we have been able to reproduce,
using additional simulations with larger softening, leading to a
central temperature decrease within a region a few times larger than
the softening radius.  The potential softening tends to produce a flat
core on the scale of the softening length.  Thus, unless we use
point-mass particles, we always see a tendency for the power index of
the density profile to approach zero at a radius comparable to the
softening radius.

After most of the merging has taken place, subsequent two-body
relaxation effects due to small $N$ continue to lead to spurious
changes in the central region. For example, a density cusp might
evolve to a flat core through gravothermal expansion (Quinlan 1996,
Fukushige and Makino 1996).  In our simulation, the local two-body
relaxation times at 1 and 5 kpc are $1.2\times 10^{10}$ and $2.3\times
10^{11}$ years, respectively.  Therefore, only the structure within
1kpc is somewhat affected by two-body relaxation in our simulation.
In simulations with $N\sim 10,000$ such as reported by Dubinski and
Carlberg (1991) and Navarro et al. (1996a, 1996b), the local
relaxation time at 5kpc was $\sim 3.0\times 10^9$ years, implying that
the effect of two-body relaxation completely dominated their results
at that distance. 

\section{Idealized Merger Simulations}

In order to illustrate the formation of temperature inversion after
merging, we performed an idealized simulation in which we merged two
equal-mass spherical halos with different central densities, a Plummer
model and a King model with central potential $W_0=9$. The result is
shown in figure 3.  The central temperature inversion is striking.

The encounter between the halos is head-on, starting from rest at an
initial separation of $R=5$ (in standard units with $M=G=-4E=1$ where
$G$ is the gravitational constant, and $M$ and $E$ are the total mass
and the initial total energy of a halo; cf. Heggie and Mathieu 1986).
The total number of particles is $N=262,144$ and the softening length
is 1/256. We used Barnes-Hut tree code on GRAPE-4 (Makino 1991b) for the
force calculations. The timestep is shared and constant ($\Delta
t=1/512$). 

In hierarchical clustering, temperature inversion takes place in the
inner halos in a similar way.  Figure 4 shows the evolution of
temperature structure for the simulation of halo formation presented
in section 3. We can see that the temperature of the outer halo
regions increases faster than the temperature of the central halo
regions. 

\section{Discussion}

In this letter, we have shown that dark halos formed through
hierarchical clustering have a central density cusp shallower than
$\rho\propto r^{-2}$, and a velocity dispersion that has a local
minimum in the center, peaks at a distance of $5\sim10$ kpc from the
center, and then drops again in the outer halo.  We offer the
following interpretations, for the temperature and the density
structures found.

The occurrence of the striking temperature inversion can be understood
as follows.  In bottom-up structure formation, as in CDM hierarchical
clustering, a typical halo is formed through repeated merging of
smaller subclumps.  Each time this happens, the less concentrated
clump tends to be disrupted by the tidal field of the more centrally
concentrated clump.  The dense core of the latter survives the merging
process more or less intact, and settles down at the center of merger
remnant, with locally unchanged temperature. In contrast, the
temperature of the merger remnant as a whole increases since the
specific binding energy of the merger remnant is almost always larger
than that of its progenitors.

As a result, present-day halos tend to carry some memory of the
temperature of their densest progenitor clump, which has set the
temperature scale in the inner, and densest, regions.  Since the bulk
of the halo is affected more by subsequent mixing and heating, a
central temperature inversion is naturally created.  This formation
process is evident, not only in our hierarchical clustering
simulations, but also in our idealized simulation of a merger between
two different halos, one with dense inner region, and another one with
a much flatter core.

The occurrence of steep density profiles, significantly steeper than
$\rho\propto r^{-1}$, naturally follows from the same physical
picture.  According to our detailed $N$-body calculations, in a purely
CDM scenario, each final halo forms a repository for the small inner
cores of the subclumps that have made up the final dark matter aggregate.

Comparing our results with observations, we conclude that the shallow
cusps of large ellipticals cannot have formed through the
dissipationless processes that we have modeled here.  Instead, we
interpret the presence of density profiles shallower than $\rho\propto
r^{-1}$ as strong evidence for the existence of central massive black
holes, through the following mechanism.  As an aftereffect of the
merger of two black-hole containing galaxies, the spiral-in of those
two massive black holes tends to smear out the central cusp that would
have otherwise formed, as shown in detailed calculations of
hierarchical merging of galaxies by Makino and Ebisuzaki (1996).

\section*{Acknowledgement}

We are grateful to Makoto Taiji, for his important contribution in
developing the GRAPE-4, and to the referee, Piet Hut, for many helpful
comments on the manuscript.  T.F.  acknowledges financial support by
JSPS.  To generate initial condition, we used the COSMIC package
developed by Edmund Bertschinger, to whom we express our thanks.  This
research was partially supported by the Grand-in-Aid for Specially
Promoted Research (04102002) of The Ministry of Education, Science and
Culture. 

\section*{References}

\medskip \noindent
Dubinski, J., end Carlberg, R., 1991, ApJ, 378, 496

\medskip \noindent
Fukushige, T., and Makino, J., 1996, in preparation. 

\medskip \noindent
Heggie, D. C., and Mathieu, R. D., 1986, in The Use of 
Supercomputer in Stellar Dynamics (ed. Hut, P., McMillan, S.) 233, 
(Springer-Verlag, Berlin)

\medskip \noindent
Hernquist, L., 1990, ApJ, 356, 359

\medskip \noindent
Lauer, T. R. et al., 1995, AJ, 110, 2622

\medskip \noindent
Makino J. 1991a, PASJ, 43, 859

\medskip \noindent
Makino J. 1991b, PASJ, 43, 621

\medskip \noindent
Makino J. and Aarseth, S. J. 1992, PASJ, 44, 141

\medskip \noindent
Makino, J., and Ebisuzaki, T., 1996, ApJ, 465, 527

\medskip \noindent
Makino, J., Taiji, M,  Ebisuzaki, T. and Sugimoto, D., 1996, submitted 
to ApJ.

\medskip \noindent
McMillan, S. L. W., 1986, in The Use of Supercomputers
in Stellar Dynamics, eds. S. L. W. McMillan, P. Hut (Springer, Berlin) p 156
 
\medskip \noindent
Merritt, D., and Fridman, T., 1996, in Fresh Views of Elliptical
Galaxies, eds. Buzzoni, A., Renzini, A. Serrano, A. (ASP conference
Series Vol. 86)

\medskip \noindent
Moore, B., 1994, Nature, 370, 629

\medskip \noindent
Navarro, J. F., Frenk, C. S., and White, S. D. M., 1996a, ApJ, 462, 563

\medskip \noindent
Navarro, J. F., Frenk, C. S., and White, S. D. M., 1996b, preprint (astro-ph/9611107)

\medskip \noindent
Quinlan, G, D., 1996, preprint (astro-ph/9606182)

\medskip \noindent
Steinmetz, M. H., and White, S. D. M., 1996, preprint (astro-ph/9609021)

\medskip \noindent
Taiji, M., Makino, J., Fukushige, T., Ebisuzaki, T., and 
Sugimoto, D., 1996, in IAU Symposium 174, eds.  Makino, J., Hut, P. 
(Kluwer Academic Press)

\medskip \noindent 
White, S. D. M. 1996, in Gravitational Dynamics, ed. Lahav, O., 
Terlevich, E., Terlevich, R. J., 121 (Cambridge University Press)

\newpage
\begin{figure}
%\epsfile{file=figure1a.ps,width=16cm}
\caption{Snapshots of $N$-body simulation of hierarchical clustering
at redshifts (a) $z=8.7$, (b)$5.1$, and (c) $1.8$.  The unit
of length  is 1 kilo parsec.}
\end{figure}
\newpage

\begin{figure}
{\centering \leavevmode \epsfxsize=18cm \epsfbox{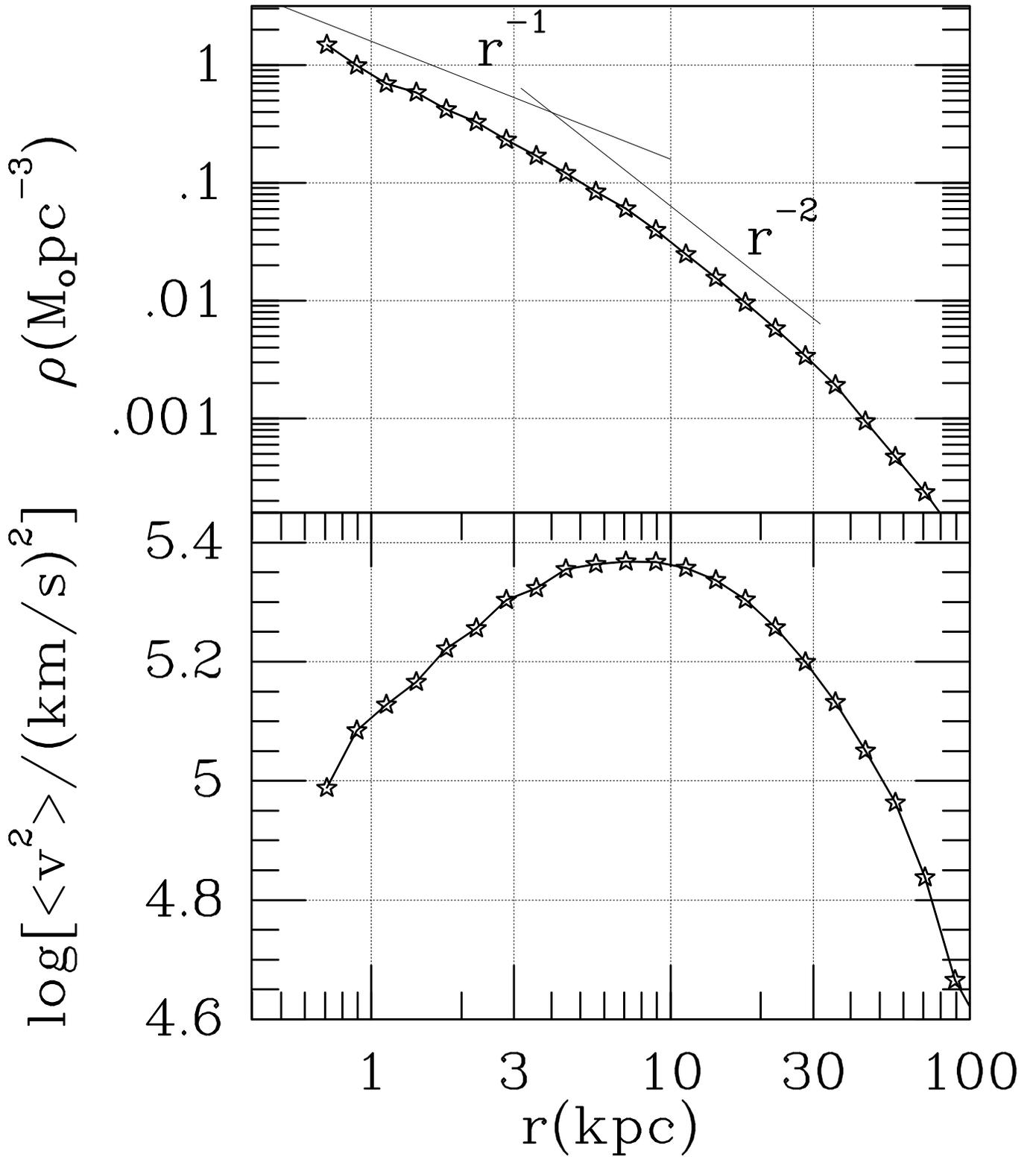}
\caption{Density and temperature structures of the halo at $z=1.8$ in
the $N$-body simulation shown in Figure 1. The position of the center
of the halo was determined using potential minimum and averaged
physical values over each spherical shell. }
}
\end{figure}

\begin{figure}
{\centering \leavevmode \epsfxsize=16cm \epsfbox{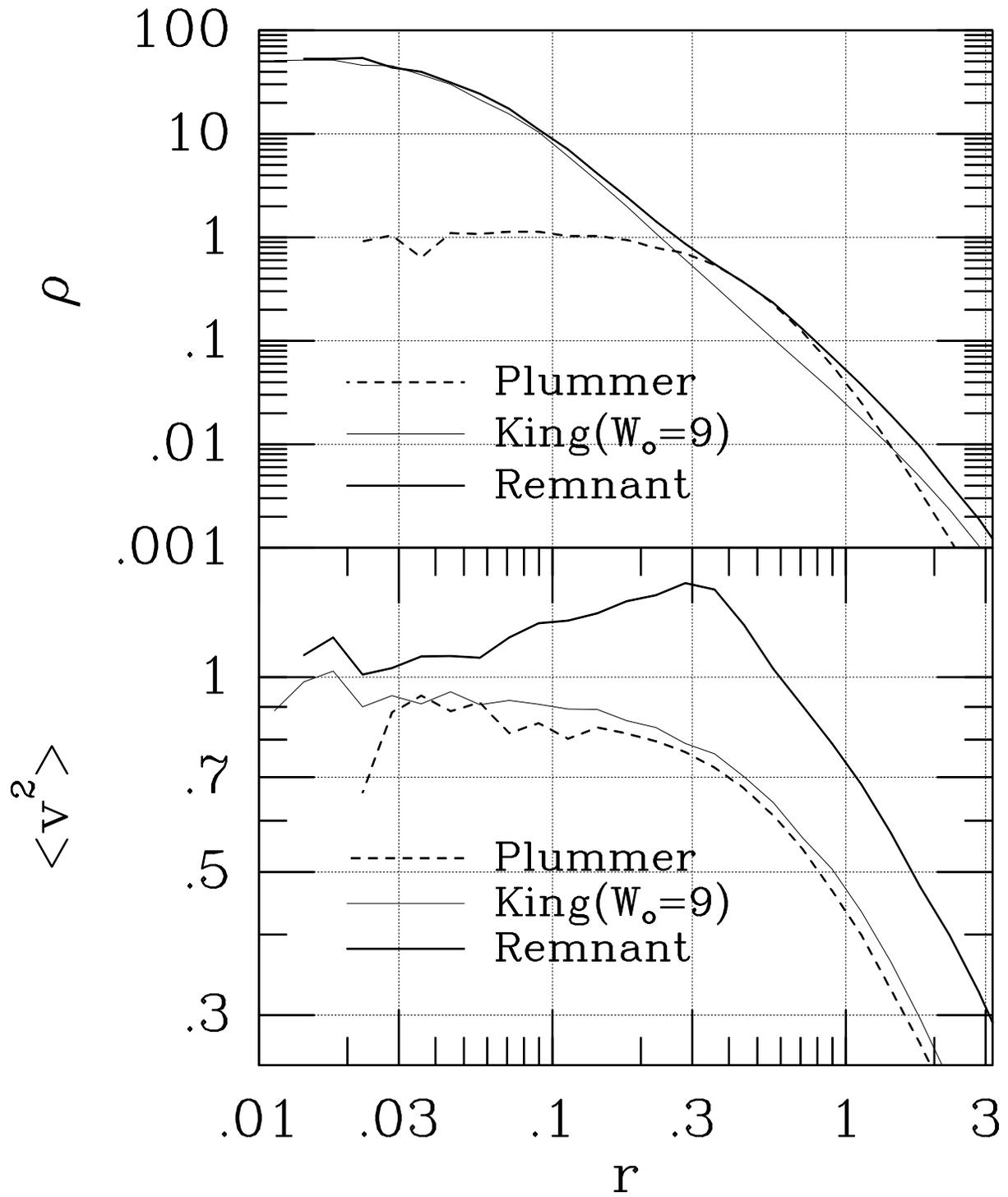}
\caption{The density and temperature structures of the merger
remnant formed by the merging of two equal-mass clusters with
different central densities.  }
}
\end{figure}

\begin{figure}
{\centering \leavevmode \epsfxsize=16cm \epsfbox{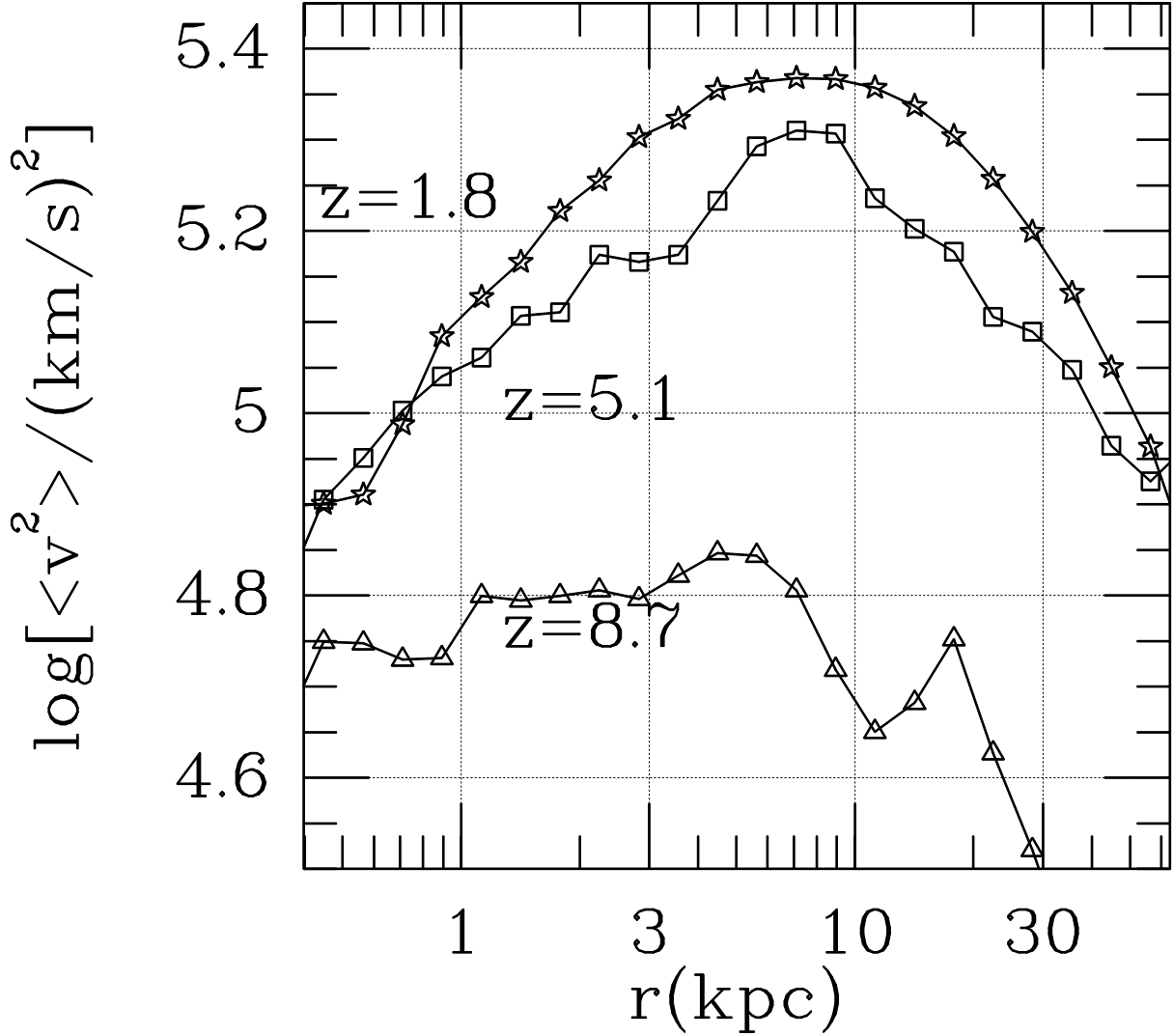}
\caption{Evolution of temperature structures of the halo at
$z=8.7$(triangle), $5.1$(square), and $1.8$(star) obtained by $N$-body
simulation illustrated in Figure 1. We made the temperature profile in
the same way as in figure 2.}
}
\end{figure}

\end{document}